# Formal Concept Analysis and Resolution in Algebraic Domains

Pascal Hitzler and Matthias Wendt

Artificial Intelligence Institute, Department of Computer Science
Dresden University of Technology, Dresden, Germany
{phitzler,mw177754}@inf.tu-dresden.de

**Abstract.** We relate two formerly independent areas: Formal concept analysis and logic of domains. We will establish a correspondene between contextual attribute logic on formal contexts resp. concept lattices and a clausal logic on coherent algebraic cpos. We show how to identify the notion of formal concept in the domain theoretic setting. In particular, we show that a special instance of the resolution rule from the domain logic coincides with the concept closure operator from formal concept analysis. The results shed light on the use of contexts and domains for knowledge representation and reasoning purposes.

## 1  Introduction

Domain theory was introduced in the 1970s by Scott as a foundation for programming semantics. It provides an abstract model of computation using order structures and topology, and has grown into a respected field on the borderline between Mathematics and Computer Science [1]. Relationships between domain theory and logic were noted early on by Scott [2], and subsequently developed by many authors, including Smyth [3], Abramsky [4], and Zhang [5]. There has been much work on the use of domain logics as logics of types and of program correctness, with a focus on functional and imperative languages.

However, there has been only little work relating domain theory to logical aspects of knowledge representation and reasoning in artificial intelligence. Two exceptions were the application of methods from quantitative domain theory to the semantic analysis of logic programming paradigms studied by Hitzler and Seda [6,7,8], and the work of Rounds and Zhang on the use of domain logics for disjunctive logic programming and default reasoning [9,10,11]. The latter authors developed a notion of clausal logic in coherent algebraic domains, for convenience henceforth called *logic RZ*, based on considerations concerning the Smyth powerdomain, and extended it to a disjunctive logic programming paradigm [11]. A notion of default negation, in the spirit of answer set programming [12] and Reiter's default logic [13], was also added [14].

The notion of *formal concept* evolved out of the philosophical theory of concepts. Wille [15] proposed the main ideas which lead to the development of formal concept analysis as a mathematical field [16]. The underlying philosophical rationale is that a concept is determined by its *extent*, i.e. the collection of objects which fall under this concept, and its *intent*, i.e. the collection of properties or attributes covered by this concept. Thus, a formal concept is usually distilled out of an incidence relation, called a *formal context*, between a set of objects and a set of attributes via some concept closure operator, see Section 2 for details. The set of all concepts is then a complete lattice under some natural order, called a *concept lattice*.

The concept closure operator can naturally be represented by an implicational theory of attributes, e.g. the attribute "is a dog" would imply the attribute "is a mammal", to give a simple example. Thus, contexts and concepts determine logical structures, which are investigated e.g. in [17,18,19]. In this paper, we establish a close relationship between the logical consequence relation in the logic RZ and the construction of concepts from contexts via the mentioned concept closure operator. We will show that finite contexts can be mapped naturally to certain partial orders such that the concept closure operator coincides with a special instance of a resolution rule in the logic RZ, and that the concept lattice of the given context arises as a certain set of logically closed theories. Conversely, we will see how the logic RZ on finite pointed posets finds a natural representation as a context. Finally, we will also see how the contextual attribute logic due to Ganter and Wille [17] reappears in our setting.

Due to the natural capabilities of contexts and concepts for knowledge representation, and the studies by Rounds and Zhang on the relevance of the logic RZ for reasoning mentioned above, the result shows the potential of using domain logics for knowledge representation and reasoning. As such, the paper is part of our investigations concerning the use of domain theory in artificial intelligence, where domains shall be used for knowledge representation, and domain logic for reasoning. The contribution of this paper is on the knowledge representation aspect, more precisely on using domains for representing knowledge which is implicit in formal contexts. Aspects of reasoning, building on the clausal logic of Rounds and Zhang and its extensions, as mentioned above, are being pursued and will be presented elsewhere, and some general considerations can be found in the conclusions. We also note that our results may make way for the use of formal concept analysis for domain-theoretic program analysis, and this issue is also to be taken up elsewhere.

The plan of the paper is as follows. In Section 2 we provide preliminaries and notation from lattice theory, formal concept analysis, and domain theory, which will be needed throughout the paper. In Section 3, we will identify certain logically closed theories from the logic RZ, called *singleton-generated theories*, and show that the set of all these coincides with the Dedekind-MacNeille completion of the underlying finite poset. This sets the stage for the central Section 4 where we will present the main results on the correspondence between concept closure and logical consequence in the logic RZ, as mentioned above. In Section 5 we shortly exhibit how the contextual attribute logic relates to our setting. Finally, in Section 6, we conclude with a general discussion on knowledge representation and reasoning perspectives of our work, and display some of the difficulties involved in carrying over the results to the infinite case.

*Acknowledgements.* We thank Sergej Kuznetsov and Guo-Qiang Zhang for inspiring discussions, and Pawel Waszkiewicz for valuable feedback on an earlier version of this paper. We are very grateful for detailed discussions with Bernhard Ganter about our work and for the comments of three anonymous referees, which helped to improve our presentation substantially.

## 2 Preliminaries and Notation

Our general reference for lattice theory is [20], assuming basic knowledge about partially ordered sets (posets) and (complete) lattices. For a poset $P$, $\uparrow X = \{y \mid x \leq y \text{ for some } x \in X\}$ denotes the *upper closure* of $X$, $X^\uparrow = \{y \mid x \leq y \text{ for all } x \in X\}$ denotes the (set of) *upper bounds* of $X$. The set of *minimal upper bounds* of a set $X$ is denoted by $\operatorname{mub} X$. The notions of *lower closure* and *lower bounds* are obtained dually.

We furthermore assume basic terminology from formal concept analysis, such as formal context, the derivation operators, and concept lattices, the standard reference being [16]. For reference, we recall a part of the basic theorem of concept lattices [16, Theorem 3], which we will use in the sequel.

**Theorem 1.** *Given a formal context $(G, M, I)$, a complete lattice $L$ is isomorphic to the concept lattice $\underline{\mathfrak{B}}(G, M, I)$ if and only if there are mappings $\overline{\gamma} : G \to L$ and $\overline{\mu} : M \to L$ such that $\overline{\gamma}(G)$ is join-dense and $\overline{\mu}(M)$ is meet-dense in $L$ and $gIm$ is equivalent to $\overline{\gamma}(g) \leq \overline{\mu}(m)$.*

Given a poset $P$, the smallest complete lattice into which $P$ can be embedded is called the *Dedekind-MacNeille completion* of $P$, which can in turn be identified with the concept lattice of the formal context $(P, P, \leq)$.

We will now recall in more detail some basic notions from domain theory and the ideas underlying domain logics, good references being [1,4,5].

We call a subset $X \subseteq P$ of a poset $P$ *directed* if for all $x, y \in X$ there exists $z \in X$ with $x \leq z$ and $y \leq z$. A poset $P$ is called *pointed* if it has a least element $\bot$, and is called a *cpo* (*complete partial order*) if it is pointed and limits of all directed subsets exist. An element $c \in P$ is called *compact* (or finite) if for all directed $D \subseteq P$ with $c \leq \bigvee D$, there exists $x \in D$ with $c \leq x$. The set of all compact elements of $P$ is denoted by $K(P)$. A cpo is called *algebraic* if every element is the directed join of the compact elements below it. A subset $U \subseteq P$ of a cpo is called *Scott open* if $\uparrow U = U$ and for any directed $D \subseteq P$ we have $\bigvee D \in U$ if and only if $U \cap D \neq \emptyset$. These sets form the so-called *Scott topology*. A cpo is called *coherent* if the intersection of any two compact open sets is compact open. A set $U \subseteq D$ is called *saturated* if it is the intersection of all Scott opens containing it. In the following, we will call coherent algebraic cpos *domains*.

In [11], Rounds and Zhang developed a clausal logic on domains, in the following called *logic RZ*, which bears potential for establishing a disjunctive logic programming paradigm with a clear domain-theoretic semantics. The next definition shortly recaptures their work. Our discussion, however, will mostly be restricted to the finite case of finite pointed posets. A short discussion of this is deferred to Section 6.

**Definition 1.** *Let $(P, \leq)$ be a domain. A* clause *over $P$ is a finite subset of $K(P)$. A* theory *over $P$ is a set of clauses over $P$. For a clause $X$ and $m \in P$, we say that $m$ is a* model *of $X$, written $m \models X$, if there exists some $x \in X$ with $x \leq m$. For a theory $T$ and $m \in P$, we set $m \models T$ if $m \models X$ for all $X \in T$, in which case we call $m$ a* model *of $T$. For a theory $T$ and a clause $X$ we say that $X$ is a* logical consequence *of $T$, written $T \models X$, if for all $m \in P$ we have that $m \models T$ implies $m \models X$. A theory $T$ is said to be* logically closed *if $T \models X$ implies $X \in T$ for all clauses $X$. Given a theory $S$, we say that $T$ is the* logical closure *of $S$ if $T$ is the smallest logically closed theory containing $S$. A theory is called* consistent *if it does not have the empty clause as a logical consequence.*

For a theory $T$, we will denote the set of models of $T$ by $\mathrm{Mod}(T)$. Similarly, given a set $M \subseteq P$ of models, we define the corresponding theory $\mathrm{Th}(M)$ to be the set of all clauses which have all elements of $M$ as model. Note that for every $M \subseteq P$ the corresponding theory $\mathrm{Th}(M)$ contains the clause $\{\bot\}$, hence is non-empty. The original rationale for studying the logic RZ was to obtain a characterization of the Smyth powerdomain of coherent algebraic cpos by means of a domain logic. The Smyth powerdomain is used in denotational semantics for modelling nondeterminism, and it can be characterized as the set of all nonemtpy compact saturated subsets of $P$, ordered by reverse subset inclusion. For details we refer to [1]. The mentioned characterization from [11] now is that the Smyth powerdomain of $P$ is isomorphic to the set of all consistent and logically closed theories over $P$, under subset inclusion.

## 3   Singleton-Generated Theories and Poset Completion

In this section, we will show a strong relationship between the logic RZ and poset completion. In particular we show that a set of certain theories is isomorphic to the Dedekind-MacNeille completion of the given poset. Due to the strong link between concept lattices and the Dedekind-MacNeille completion exhibited by Theorem 1, this will provide the necessary tool for our main results, presented in Section 4.

First of all, we describe the domain-theoretic Smyth powerdomain construction by means of formal concept analysis. Note however, that although the Smyth powerdomain of a coherent algebraic cpo is always a lattice (just missing a top element), it is in general not a complete lattice. Thus the assumption of $P$ being a finite pointed poset is crucial to our formal concept analysis reformulation of this powerdomain construction. Now in the finite case, we can provide two representations of the Smyth powerdomain of a pointed poset $P$: The first one uses the definition of the Smyth powerdomain as the set of all compact saturated subsets of $P$ ordered by reverse inclusion. In the finite case, a set is compact saturated if and only if it is upward closed. This yields that the Smyth powerdomain of a finite pointed poset is isomorphic to $\underline{\mathfrak{B}}(P, P, \not\geq)$, having as intents the order filters, ordered by reverse inclusion. The second representation uses the logical characterization from [11] mentioned above, according to which the Smyth powerdomain of a given domain $P$ is isomorphic to the set of all logically closed consistent theories over $P$, under subset inclusion. Since clauses in the finite case are just subsets of $P$, the Smyth powerdomain is isomorphic to $\underline{\mathfrak{B}}(\mathfrak{P}(P), P, \models)$, where $\mathfrak{P}(P)$ denotes the powerset of $P$. Corresponding concepts are of the form $(A, B)$, where $A$ is a logically closed theory and $B$ is its set of models, i.e., $A = \text{Th}(B)$ and $B = \text{Mod}(A)$.

We will now investigate conjunctive assertions in the logic RZ. According to the intuition that theories are conjunctions of their clauses, we have to consider theories which contain only singleton clauses. Theories which are the closure of a set containing only singleton clauses will be called *singleton-generated theories*, and a set $\mathcal{G} \subseteq P$ will be called a *generator* of the theory $\text{Th}(\{\{d\} \mid d \in \mathcal{G}\})$. These definitions formalize the idea of considering only inferences of the form $\{\{d_1\}, \ldots, \{d_n\}\} \models \{d\}$, which was also done in [21].

These singleton-generated theories can be obtained by restricting the objects of the Smyth powerdomain context $(\mathfrak{P}(P), P, \models)$ to singleton clauses. Thus, the singleton-generated theories are obtained as the extents of the formal context $(P, P, \models)$. Noting that for a singleton clause $\{d\}$ and a model $m$ we have $\{d\} \models m$ if and only if $d \leq m$, the context for the singleton-generated theories can be written as $(P, P, \leq)$. We have just shown the following.

**Theorem 2.** *Given a finite pointed poset $P$, the set of all singleton-generated theories over $P$, ordered by subset inclusion, is isomorphic to the Dedekind-MacNeille completion $\mathcal{N}(P) \cong \underline{\mathfrak{B}}(P, P, \leq)$ of $P$.*

Before we make use of Theorem 2 in the next section, let us briefly reflect on what we have achieved so far. Identifying singleton-generated theories with elements of the Dedekind-MacNeille completion yields the possibility of representing finite lattices — which are always complete — by means of finite pointed posets. From an order-theoretic point of view this idea appears to be rather straightforward. Relating this setting to a logic of domains, however, provides a novel aspect. On the one hand, we now have the possibility to use a restricted form of resolution on ordered sets — as will be explained in Section 4 — in order to represent elements of the corresponding Dedekind-MacNeille completion. On the other hand, we obtain a new perspective on the logic RZ, namely that underlying posets can be interpreted from a knowledge representation point of view. More precisely, in the next section we will show how Theorem 2 can be employed for relating the logic RZ to formal concept analysis. The following corollary to Theorem 2 will also be helpful as it provides a logical representation of finite lattices by theories on finite pointed posets.

**Corollary 1.** *Let $L$ be a finite lattice. Then for every finite pointed poset $P$ which can be embedded join- and meet-densely into $L$, the set of all singleton-generated theories over $P$ is isomorphic to $L$.*

We know that every complete lattice is the concept lattice of some formal context. We can thus interpret the elements of the Dedekind-MacNeille completion of a pointed poset $P$, which can in turn be identified with singleton-generated theories over $P$, as concepts in the corresponding concept lattice. This indicates that the logic RZ might be used as a knowledge representation formalism. This, and details of the relationship between the logic RZ on finite pointed posets and concept lattices will be explained in more detail in the next section.

## 4   Representation of Formal Contexts by Finite Posets

In the previous section we have shown that each finite pointed poset $P$ can be interpreted as a formal context whose concept lattice $\underline{\mathfrak{B}}(P, P, \leq)$ is isomorphic to the set of all singleton-generated theories over $P$. In this formal context, each element of the poset is both object and attribute, resembling the fact that in the domain logic of a finite pointed poset, each element can be both a singleton clause and a model.

Since we consider finite lattices, [16, Proposition 12] implies that there is a unique reduced context, the standard context of $(P, P, \leq)$, having a concept lattice isomorphic to $\underline{\mathfrak{B}}(P, P, \leq)$. The objects of the standard context are those elements of $P$ whose object concept is join-irreducible in $\underline{\mathfrak{B}}(P, P, \leq)$, the attributes are those whose attribute concept is meet-irreducible. Considering the principal ideal embedding of $P$ into $\underline{\mathfrak{B}}(P, P, \leq)$, we find that the irreducible objects of $(P, P, \leq)$ are those which are not the join of all elements strictly below them, whereas the irreducible attributes are those which are not the meet of all elements strictly above them.

From these considerations it follows immediately that any singleton-generated theory over $P$ is completely determined by the set of irreducible objects it contains as singletons: any singleton which is not an irreducible object can be represented as the join of all the irreducible objects below it, hence is derivable from these objects in the logic RZ. Formally, we can state the following lemma.

**Lemma 1.** *Let $P$ be a finite pointed poset and $T_1$, $T_2$ be two singleton-generated theories over $P$ which coincide on all irreducible objects of $P$. Then $T_1 = T_2$.*

*Proof.* Assume $\{x\} \in T_1$ for some reducible object $x \in P$. Now $x$ is the join of all the irreducible objects below it, and by logical closure $\{m\} \in T_1$ for all elements $m \leq x$, hence $\{m\} \in T_2$ for all irreducible objects $m \leq x$. So by logical closure of $T_2$ we obtain $\{x\} \in T_2$. The argument clearly reverses and therefore suffices.

At this stage our way of interpreting singleton-generated theories as concepts becomes almost obvious. An object $g \in P$ is in the extent of a concept (i.e. of a singleton-generated theory) $T$ if $\{g\} \in T$, and an attribute $m \in P$ is in the intent of a concept $T$ if $m \models T$. The latter means that every object contained as singleton in $T$ necessarily has attribute $m$. Furthermore, if an object $g$ is a model for a theory, then it is necessary for any other object in the corresponding concept extent to have all the attributes $g$ has. If an attribute is contained as a singleton in a theory, then any object having this attribute is also contained as a singleton in the theory.

When reasoning about the knowledge represented in a poset $P$, we can — according to Lemma 1 — restrict our attention to the irreducible objects. But we can also incorporate the attributes into the reasoning, if desired, as a kind of *macros* for describing collections of objects. This perspective will be employed later on when discussing the logic programming framework developed by Rounds and Zhang [11] in terms of formal concept analysis. Reasoning with objects probably seems unusual from the point of view of formal concept analysis, since it is more common to consider the logic behind the attributes, focusing on the implicational theory of the attributes. The emphasis on objects in this paragraph stems from the domain theoretic intuition on which the logic RZ is based, namely that $x \leq y$ stands for the situation in which $y$ contains more information than $x$. When representing concept lattices by singleton generated theories in the logic RZ, it is therefore more intuitive to consider the concept lattice in its reverse order, or equivalently, the dual context in which the attributes are considered to be the new objects, with which we reason. We will do this in the following.

The next theorem makes explicit the representation of finite formal contexts by finite pointed posets, such that the concept lattice of the formal context is isomorphic (in the reverse order) to the set of all singleton generated theories of the finite pointed poset.

**Theorem 3.** *Let $(G, M, I)$ be a reduced finite formal context. Then $\underline{\mathfrak{B}}(G, M, I)$, under the reverse order, is isomorphic to the set of all singleton-generated theories of a finite pointed poset $P$, under subset inclusion, if and only if there exist bijections $\gamma : G \to M(P)$ and $\mu : M \to J(P)$, where $(J(P), M(P), \leq)$ is the reduced context of $(P, P, \leq)$, such that for all $g \in G$ and $m \in M$ we have $gIm$ if and only if $\gamma(g) \geq \mu(m)$.*

*Proof.* Immediate from [16, Proposition 12], the Basic Theorem of Concept Lattices (Theorem 1) and Theorem 2.

So the singleton-generated theories in the logic RZ have the same implicational logic as the attributes of the represented context $(G, M, I)$, whenever we restrict to the irreducible objects.

Next, we will give a specific example for a construction of a finite pointed poset from a given context $(G, M, I)$. Note that also $\underline{\mathfrak{B}}(G, M, I)$, reversely ordered, is a finite pointed poset trivially satisfying the conditions from Theorem 3.

*Example 1.* Let $(G, M, I)$ be a formal context, where $G$ and $M$ are finite and disjoint. Define the following ordering on $G \mathrel{\dot\cup} M$:

(i) For $m_1, m_2 \in M$ let $m_1 \leq m_2$ if $m_1' \supseteq m_2'$.
(ii) For $g_1, g_2 \in G$ let $g_1 \leq g_2$ if $g_1' \subseteq g_2'$.
(iii) For $g \in G$ and $m \in M$ let $m \leq g$ if $gIm$.
(iv) For $g \in G$ and $m \in M$ let $g \leq m$ if for all $h \in G$ and all $n \in M$ we have that $gIn$ and $hIm$ imply $hIn$.

The above construction yields a preorder on $G \mathrel{\dot\cup} M$. We obtain from this a partial order, also denoted by $\leq$, by taking the quotient order in the usual way.

**Table 1.** Formal context for Example 2.

|   | salad | starter | fish | meat | red wine | white wine | water | dessert | coffee | expensive |
|---|---|---|---|---|---|---|---|---|---|---|
| 1 |   |   | × |   |   | × |   | × |   |   |
| 2 |   |   |   | × | × |   |   | × |   |   |
| 3 | × |   | × |   |   | × |   | × | × | × |
| 4 |   | × |   | × | × |   |   | × | × | × |
| 5 | × | × | × |   |   |   | × |   |   |   |
| 6 | × | × |   | × |   |   | × |   | × |   |
| 7 | × | × |   | × | × |   | × | × | × | × |
| 8 |   |   |   | × |   |   | × |   | × |   |
| 9 | × | × |   |   |   |   |   |   | × |   |

If $(G \,\dot\cup\, M/_\sim, \leq)$ does not have a least element, we add $\bot$ to $G \,\dot\cup\, M/_\sim$ and set $\bot \leq x$ for all $x \in G \,\dot\cup\, M/_\sim$. The latter amounts to adding an additional attribute $m$ with $m' = G$ to the context.

The main intuition behind this construction is to use the set consisting of all objects and attributes as a join- and meet-dense subset of the concept lattice and to supply the induced order by constructions directly available from the formal context. The first three items do exactly this. However, we have to take care of those elements which are both join- and meet-irreducible in the concept lattice. This is achieved with (iv) and the quotient order construction, where those object-attribute pairs are identified which will result in doubly irreducible elements. This construction of endowing $G \,\dot\cup\, M/_\sim$ with the induced order from $\underline{\mathfrak{B}}(G, M, I)$ is also known as the Galois subhierarchy introduced in [22], see also [23] and the references contained therein. Formally, we can state the following.

**Proposition 1.** *Given a finite formal context $(G, M, I)$ and the poset $P = (G \,\dot\cup\, M/_\sim, \leq)$ as defined in Example 1, we have $\underline{\mathfrak{B}}(G, M, I) \cong \underline{\mathfrak{B}}(P, P, \leq)$.*

*Example 2.* Consider the formal context given in Table 1. It shall represent, in simplified form, a selection of set dinners from a restaurant menu. Using Example 1, we obtain the finite pointed poset depicted in Figure 1. Concepts in this setting correspond to types of dinners, e.g. one may want to identify the concept with extent $\{4, 6, 7\}$ and intent $\{st, m, c\}$, using the abbreviations from Figure 1, to be the *heavy* meals, while the *expensive* ones are represented by the attribute concept of $e$, and turn out to always include coffee. Using the logic RZ, we can for example conclude that a customer who wants salad and fish will choose one of the meals 3 or 5, since these elements of the poset are exactly those which are both objects and models of the theory $\{\{sd\}, \{f\}\}$. Also, he will always get a starter or a dessert, formally $\{\{sd\}, \{f\}\} \models \{st, d\}$. To give a slightly more sophisticated example, suppose that a customer wants salad or a starter, additionally fish or a dessert, and drinks water. From this we can conclude that in any case he will get both a salad and a starter. Formally, we obtain $\{\{sd, st\}, \{f, d\}, \{w\}\} \models \{sd\}$ and $\{\{sd, st\}, \{f, d\}, \{w\}\} \models \{st\}$. A little bit of reflection on the context makes it clear that these inferences are indeed natural ones, in fact also follow from the implicational theory of the context.

**Fig. 1.** Figure for Example 2. Abbreviations are: $sd$ salad, $st$ starter, $f$ fish, $m$ meat, $rw$ red wine, $ww$ white wine, $w$ water, $d$ dessert, $c$ coffee, $e$ expensive.

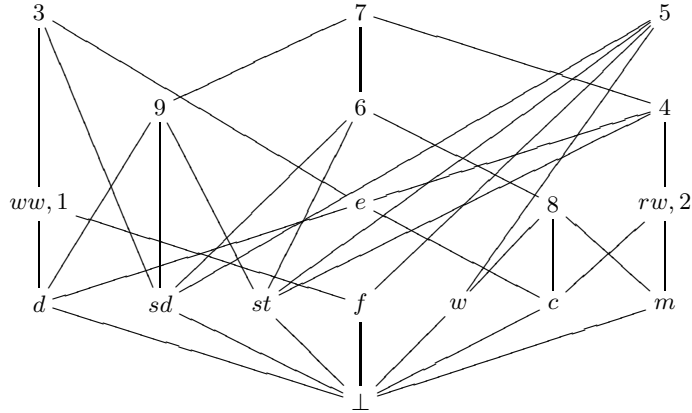

Having established Theorem 3 as a link between the logic RZ and formal concept analysis, we will now discuss how the different techniques on both sides embed. In particular, we will shortly consider a proof theory for the logic RZ, discussed next, and also the contextual attribute logic of formal concept analysis, discussed in Section 5.

In [11], the following hyperresolution rule was presented:

$$\frac{X_1\ X_2 \ldots X_n;\quad a_i \in X_i \quad \text{for } 1 \leq i \leq n;\quad \text{mub}\{a_1, \ldots, a_n\} \models Y}{Y \cup \bigcup_{1 \leq i \leq n}(X_i \setminus \{a_i\})}$$

In words, this rule says that from clauses $X_1, \ldots, X_n$, $a_i \in X_i$ for all $i$, and $\text{mub}\{a_1, \ldots, a_n\} \models Y$ with respect to the logic RZ, the clause $Y \cup \bigcup_{1 \leq i \leq n}(X_i \setminus \{a_i\})$ may be derived. This rule, together with two special rules treating the cases of an empty selection of clauses resp. an empty clause in the premise of the rule, yields a proof theory resp. entailment relation $\vdash$ which is sound and complete w.r.t. the model theory given in Definition 1. From our results, in particular from Theorem 3, we obtain that the following restriction of the hyperresolution rule to singleton clauses induces an entailment relation $\vdash_s$ which is equivalent to the concept closure operator $(\cdot)'' : \mathfrak{P}(G) \to \mathfrak{P}(G)$ which maps any set of objects $B$ to the extent $B''$ of the corresponding concept $(B'', B')$:

$$\frac{\{a_1\}\ \{a_2\}\ldots\{a_n\};\quad \mathrm{mub}\{a_1,\ldots,a_n\}\models\{a\}}{\{a\}}$$

Thus we can conclude that the logic RZ can be used for knowledge representation in much the same way as formal concept analysis: There is a correspondence between finite formal contexts and finite pointed posets and, moreover, both the proof and the model theory of [11] lend themselves to an easy characterization of concept closure. This is probably not too surprising from the viewpoint of formal concept analysis resp. lattice theory. However, from the viewpoint of domain theory it is certainly interesting that there is such a close correspondence between domain logics developed for reasoning about program semantics [4] and a knowledge representation mechanism like formal concept analysis.

## 5  Contextual Attribute Logic and the Logic RZ

In this section, we will show that the correspondence between the logic RZ and formal concept analysis is not exhausted by the relationship between singleton-generated theories and concept closure. In particular, we will show how to identify part of the contextual attribute logic due to [17] in a finite pointed poset $P$ by means of the logic RZ. We first show how clauses and theories resemble constructions of compound attributes in the poset.

In [17], compound attributes are defined to be compositions of attributes w.r.t. their extent. More precisely, for any set $A \subseteq M$ of attributes of a formal context $(G, M, I)$, the compound attribute $\bigvee A$ has the extent $\bigcup\{m' \mid m \in A\}$, and the compound attribute $\bigwedge A$ has the extent $\bigcap\{m' \mid m \in A\}$. For an attribute $m \in M$, the compound attribute $\neg m$ has the extent $G \setminus m'$.

Now we can relate compound attributes and theories in the logic RZ by the following proposition, which is in fact a straightforward consequence of our previous results, so we skip the proof.

**Proposition 2.** *Let $P$ be a finite pointed poset and consider the formal context $(G, M, I)$ obtained from it as indicated in Theorem 3, and let $\gamma$, $\mu$ be as in the same theorem. Then for all $A \subseteq M$, $g \in G$, and $m \in M$ the following hold.*

- *$g$ is in the extent of $\bigvee A$ if and only if $\gamma(g) \models \mu(A)$.*
- *$g$ is in the extent of $\bigwedge A$ if and only if $\gamma(g) \models \{\{\mu(a)\} \mid a \in A\}$.*
- *$g$ is in the extent of $\neg m$ if and only if $\gamma(g) \not\models \{\mu(m)\}$.*

We thus see, that the formation of conjunction and disjunction of attributes to compound attributes corresponds exactly to the formation of singleton generated theories resp. clauses. Negation, however, is more difficult to represent in the logic RZ, since the set of all models of $\neg m$ is not an upper set, but a lower set, more precisely it is the complement of a principal filter in $P$. Thus it seems that the Scott topology, on which the logic RZ is implicitly based, is not appropriate for handling this kind of negation — which could be a candiate for a strong negation in the logic programming paradigm discussed in Section 6. It remains to be investigated whether the results presented in [11] carry over to the Lawson topology and the Plotkin powerdomain, see [1] for definitions, which according to what has been said above may be the correct setting for handling this negation.

In [17], sequents of the form $(A, S)$, where $A, S \subseteq M$, were introduced as a possible reading of compound attributes $\bigvee(S \cup \{\neg m \mid m \in A\})$. A sequent $(A, S)$ may thus be interpreted as an implication $\bigvee S \leftarrow \bigwedge A$. A *clause set* over $M$ is a set of sequents over $M$. The clause logic, called *contextual attribute logic*, of a finite pointed poset $P$ is then the set of all sequents that are *all-extensional* in $P$, i.e. all sequents whose extent contains the set of all objects of $P$. This means that the implication represented by the sequent holds for all the objects in $P$.

Due to the difficulties with negation discussed above we restrict our attention, for the time being, to all-extensional sequents $(A, S)$ with $A = \emptyset$. So consider again the setting of Proposition 2, and let $X \subseteq M$. Then $(\emptyset, X)$ is an all-extensional sequent if and only if $\gamma(x) \models \mu(X)$ for all $x \in G$. This is easily verified using Theorem 3 and Proposition 2.

Apart from investigating compound attributes involving negation — as discussed above — it also remains to be determined whether there exists a way of identifying the contextual attribute logic by means of the proof theory defined by Rounds and Zhang [11]. This will be subject to further research.

## 6  Conclusions and Further Work

We have displayed a strong relationship between formal concept analysis and the domain logic RZ. The restriction of inference to singleton clauses yields the concept closure operator of formal concept analysis. Furthermore, any logically closed theory in the logic RZ can be understood as a clause set over a formal context, in the sense of contextual attribute logic, and the hyperresolution rule of [11] can be used to reason about such knowledge present in a given formal context in much the same way as the resolution rule proposed in [17]. This of course can be a foundation for logic programming over formal contexts, i.e. logic programming with background knowledge which is taken from a formal context and used as "hard constraints".

The appropriate way of doing this on domains was also studied by Rounds and Zhang. In their logic programming paradigm on coherent algebraic cpos a logic program is a set of rules of the form $\theta \leftarrow \tau$, where $\theta$ and $\tau$ are clauses over the respective domain. The models of such a rule $\theta \leftarrow \tau$ are exactly those elements $w$ of the domain which satisfy $w \models \theta$ whenever $w \models \tau$. The rule

$$\frac{X_1\ X_2 \ldots X_n;\quad a_i \in X_i \quad \text{for } 1 \leq i \leq n;\quad \theta \leftarrow \tau \in P;\quad \text{mub}\{a_1, \ldots, a_n\} \models \tau}{\theta \cup \bigcup_{1 \leq i \leq n}(X_i \setminus \{a_i\})}$$

corresponds to inference taking the clause $\theta \leftarrow \tau$ into account. By adjoining to the usual proof theory the inference rules for all the clauses in a given program, one can define a monotonic and continuous operator $T_P$ on the set of all logically closed theories, whose least fixed point yields a very satisfactory semantics for the considered program $P$.

This logic programming paradigm can be understood as logic programming with background knowledge, since the semantics of the program does not only satisfy the program, in a reasonable sense, but also takes into account those implications which are hidden in the underlying domain, i.e. in the context. It is interesting to note that the knowledge implicit in the context need not be made explicit, e.g. by computing the stem base of the context. This implicational knowledge is implicitly represented by the inference rules constituting the proof theory of the logic RZ. The authors are currently investigating the potential of this approach.

*Example 3.* Consider again the setting from Example 2 and suppose that a customer's wishes can be expressed by the following rules.

$$\{rw\} \leftarrow \{m\}$$
$$\{c, ww\} \leftarrow \{d\}$$
$$\{sd, st\} \leftarrow \{\bot\}$$

We understand, that the customer does not want meat without red wine, that for him a dessert should always go with white wine or coffee, and that in any case he wants a starter or a salad. The models for this program are 3, 4, 5, 7, $sd$, and $st$, which do not constitute an upper set. Possible choices by the customer are again those models which are also objects, i.e. 3, 4, 5, and 7. Drawing inferences from this program alone, however, yields counterintuitive results, e.g. because every clause which is a logical consequence must contain both $st$ and $sd$ (or the bottom element). Consequently, $\{e, f\}$ is not a logical consequence although each possible choice of meal by the customer will include fish or be expensive. This situation is caused by the fact that objects and attributes are no longer distinguished in the domain-theoretic setting, and that the information that salad or starter are part of every meal which is satisfactory for the customer, is necessarily part of every inference drawn. We can rectify this by adding another rule

$$\{1, 2, 3, 4, 5, 6, 7, 8, 9\} \leftarrow \{\bot\}$$

to the program, which can be interpreted as saying that the customer also wants a meal (obviously). The logical consequences from this program are then as can be expected from the context, e.g. $\{e, f\}$ as discussed above. So evaluating the set of rules amounts to querying the background knowledge represented by the context from Table 1. We suspect a strong relationship to inferences from the contextual attribute logic underlying the context, but details remain to be worked out.

In this paper, we have restricted our considerations to the case of finite pointed posets. So let us shortly discuss some of the difficulties involved in carrying over our results to the case of arbitrary coherent algebraic cpos. The correspondence between singleton-generated theories and Dedekind cuts underlying

Theorem 2 carries over to the infinite case without major restrictions — one just has to correctly adjust it to compact elements and to keep in mind that any non-compact element can be represented as the supremum of all compact elements below it. Difficulties occur when trying to characterize the lattices which arise as Dedekind-MacNeille completions of coherent algebraic cpos, since on the domain-theoretic side one has to deal with the topological notion of coherence, which is not really present on the lattice-theoretic side. Furthermore, the Scott topology we are implicitly dealing with when working with the logic RZ is not completion invariant, which means that the properties defined in terms of the Scott topology, e.g. continuity of the poset, do not carry over to the completion [24]. These issues will also have to be subject to further research. A construction similar to Example 1 carries over to a restricted infinite case, and details can be found in [21].

We finally note the very recent paper by Zhang [25], which also studies relationships between domain theory and formal concept analysis, though from a very different perspective involving Chu spaces.